\documentclass[12pt,preprint]{aastex}

\def\subsun{\mbox{$_{\odot}$}}
\def\lesssim{\mathrel{\hbox{\rlap{\hbox{%
 \lower4pt\hbox{$\sim$}}}\hbox{$<$}}}}
\def\gtrsim{\mathrel{\hbox{\rlap{\hbox{%
 \lower4pt\hbox{$\sim$}}}\hbox{$>$}}}}

\shorttitle{First to Second Stars}
\shortauthors{Smith \& Sigurdsson}

\begin{document}

\title{The Transition from the First Stars to the Second Stars in the Early Universe} 

\shorttitle{First to Second Stars}

\author{Britton D. Smith and Steinn Sigurdsson}
\affil{525 Davey Laboratory, 
  Department of Astronomy \& Astrophysics, 
  The Pennsylvania State University, 
  University Park, PA, 16802}
\email{britton@astro.psu.edu,steinn@astro.psu.edu}

\begin{abstract}
  We observe a sharp transition from a singular, high-mass mode of star formation, to 
  a low-mass dominated mode, in numerical simulations, at a metallicity of 10$^{-3}$ Z$\subsun$.  
  We incorporate a new method for including the radiative 
  cooling from metals into adaptive mesh-refinement hydrodynamic simulations.  
  Our results illustrate how metals, produced by the first stars, led to a 
  transition from the high-mass star formation mode of Pop III stars, to the low-mass mode 
  that dominates today.  We ran hydrodynamic simulations with cosmological initial 
  conditions in the standard $\Lambda$CDM model, with 
  metallicities, from zero to 10$^{-2}$ Z$\subsun$, 
  beginnning at redshift, z = 99. The simulations were run until a dense core 
  forms at the center of a 5 $\times$ 10$^{5}$ M$\subsun$ dark matter halo, at z 
  $\sim$ 18.  Analysis of the central 1 M$\subsun$ core reveals that 
  the two simulations with the lowest metallicities, Z = 0 and 10$^{-4}$ Z$\subsun$, 
  contain one clump with 99\% of the mass, while the two with 
  metallicities, Z = 10$^{-3}$ and 10$^{-2}$ Z$\subsun$, each contain two clumps that share 
  most of the mass.  The Z = 10$^{-3}$ Z$\subsun$ simulation also produced 
  two low-mass proto-stellar objects with masses between 10$^{-2}$ and 10$^{-1}$ M$\subsun$.  
  Gas with Z $\ge$ 10$^{-3}$ Z$\subsun$ is able to cool to the temperature of the 
  CMB, which sets a lower limit to the minimum fragmentation mass.  
  This suggests that the second generation stars produced a spectrum of lower mass stars, but 
  were still more massive on average than stars formed in the local universe.
\end{abstract}


\section{Introduction}

Numerical simulations have shown that the very first stars invariably formed in isolation 
and were much more massive than the sun, due mainly to the inability of primordial gas to 
efficiently cool at low temperatures 
\citep{2002Sci...295...93A,2002ApJ...564...23B,2006ApJ...652....6Y}.  
\citet{2004ApJ...612..602T} have suggested that the Pop III IMF was not dominated by 
very massive stars (M $>$ 140 M$\subsun$), but instead by stars with M = 8--40 M$\subsun$.  
Even this IMF, though, is still remarkably distinct from that observed for the local universe, 
which peaks at less than one solar mass 
\citep{1979ApJS...41..513M,2002Sci...295...82K,2003PASP..115..763C}.

The deaths of the first stars produced and distributed copious amounts of metals into their 
surroundings, through either core-collapse (M $\gtrsim$ 10 M$\subsun$) 
or pair-instability (M $\gtrsim$ 140 M$\subsun$) supernovae \citep{2002ApJ...567..532H}.  
These metals provide additional avenues for radiative cooling of the ambient gas, through 
fine-structure and molecular transitions, as well as continuum emission from dust formed from 
the supernova ejecta, permitting the gas that will form the next generation of stars to 
reach temperatures lower than what is possible for metal-free gas.  Fragmentation of collapsing 
gas will continue so long as the gas can keep decreasing in temperature as the density increases 
\citep{2005MNRAS.359..211L}, or until the gas becomes optically thick to its own emission 
\citep{1976MNRAS.176..367L}.  The minimum fragment mass is determined by the local Jeans mass, 
\begin{equation} \label{eqn:mj}
M_{J} \simeq 700 \textrm{ M}\subsun (T/200 K)^{3/2} (n/10^{4} cm^{-3})^{-1/2} (\mu/2)^{-2},
\end{equation}
where T, n, and $\mu$ are the temperature, number density, and mean molecular weight, at the 
halt of fragmentation \citep{2005MNRAS.359..211L}.  For metal-free gas, a minimum temperature 
of $\sim$ 200 K is reached at n $\simeq$ 10$^{4}$ cm$^{-3}$ when H$_{2}$ cooling becomes 
inefficient, yielding a Jeans mass, M$_{J}$ $\simeq$ 10$^{3}$ M$\subsun$ 
\citep{2002Sci...295...93A,2002ApJ...564...23B}.  At some certain chemical abundance, it is 
conjectured that metals provide sufficient cooling, so that the temperature 
of the gas continues to decrease as the density increases past the stalling point for metal-free 
gas, allowing the collapsing gas-cloud to undergo fragmentation and form smaller and smaller 
clumps.  The enrichment of gas to some critical metallicity, 
Z$_{cr}$, will trigger the formation of the first low-mass (Pop II) stars in 
the universe, as the gas can cool to lower temperatures at higher metallicity, in general.
The value of Z$_{cr}$ can be estimated by calculating the metallicity required to 
produce a cooling rate equal to the rate of adiabatic compression heating at a given 
temperature and density.  This has been carried out for individual alpha elements, 
such as C and O, by \citet{2003Natur.425..812B}, and C, O, Si, Fe, as well as solar 
abundance patterns by \citet{2006ApJ...643...26S}, yielding roughly, 10$^{-3.5}$ Z$\subsun$ 
$\lesssim$ Z$_{cr}$ $\lesssim$ 10$^{-3}$ Z$\subsun$.

Aside from the minimum clump mass, however, not much more can be said about the 
spectrum of clump masses produced during fragmentation.  \citet{2005ApJ...626..627O} 
use one-zone models with very sophisticated chemical networks to follow the evolution 
of temperature and density in the center of a collapsing gas cloud, for a range of 
metallicities.  The predictions of fragmentation from this work, though, are based 
solely on statistical arguments of elongation in prestellar cores and do not capture 
the complex processes of interaction and accretion associated with the formation of 
multiple stars \citep{2003MNRAS.339..577B}.  \citet{2006ApJ...642L..61T} simulate the 
high density (n $\ge$ 10$^{10}$ cm$^{-3}$) evolution of extremely low-metallicity 
gas (Z $<$ 10$^{-4}$ Z$\subsun$), but the conclusions of 
this work are limited by the fact that the simulations are initialized at an extremely 
late phase in the evolution of the prestellar core.  The numerical simulations by 
\citet{2001MNRAS.328..969B}, which use cosmological initial conditions, show 
fragmentation in gas with Z = 10$^{-3}$ Z$\subsun$, but a mass resolution of 
100 M$\subsun$ prevents this study from saying anything conclusive about the 
formation of sub-stellar mass objects.

In this paper, we present the results of three-dimensional hydrodynamic simulations of 
metal-enriched star-formation.  These simulations are similar in nature to those of 
\citet{2001MNRAS.328..969B}, but with vastly improved numerical methods and updated physics.  
We describe the setup of our simulations in \S\ref{sec:setup}, with the 
results in \S\ref{sec:results} and a discussion of the consequences of this work in 
\S\ref{sec:discussion}.

\section{Simulation Setup} \label{sec:setup} 

We perform a series of four simulations, with constant metallicities, Z = 0 (metal-free), 
10$^{-4}$ Z$\subsun$, 10$^{-3}$ Z$\subsun$, and 10$^{-2}$ Z$\subsun$, 
using the Eulerian adaptive mesh refinement 
hydrodynamics/N-body code, Enzo \citep{1997WSAMRGMBryan,2004CWAMROShea}.  
The metallicity is held constant throughout each simulation in order to isolate the role 
of heavy element concentration in altering the dynamics of collapse compared to the 
identical metal-free case.  In reality, metals will be injected over time into star forming 
gas by Pop III supernova blast waves, and the mixing of those metals with the gas will not 
be completely uniform. Here we focus on an idealized approximation in order to capture 
the essential physics of collapse and fragmentation.

Each simulation begins at z = 99, in a cube, 300 h$^{-1}$ kpc comoving per side, in a 
$\Lambda$CDM universe, with the following cosmological parameters: $\Omega_{M}$ = 0.3, 
$\Omega_{\Lambda}$ = 0.7, $\Omega_{B}$ = 0.04, and Hubble constant, h = 0.7, in units of 
100 km s$^{-1}$ Mpc$^{-1}$.  We initialize all the simulations identically, with a power 
spectrum of density fluctuations given by \citet{1999ApJ...511....5E}, with $\sigma_{8}$ 
= 0.9 and n = 1.  The computation box consists of a top grid, with 128$^{3}$ cells, and 
three static subgrids, refining by a factor of 2 each.  This gives the central refined 
region, which is 1/64 the total computational volume, an effective top grid resolution 
of 1024$^{3}$ cells.  The grid is centered on the location of a $\sim$ 5 $\times$ 10$^{5}$ 
M$\subsun$ dark matter halo that is observed to form at z $\sim$ 18 in a prior 
dark-matter-only simulation, as is done similarly in \citet{2002Sci...295...93A,
2005ApJ...628L...5O}.  Refinement occurs during the simulations whenever the 
gas, or dark matter, density is greater than the mean density by a factor of 4, 
or 8, respectively.  We also require that the local Jeans length be resolved by at least 
16 grid cells at all times in order to avoid artificial fragmentation as prescribed by 
\citet{1997ApJ...489L.179T}.

To include the radiative cooling processes from the heavy elements, we use the method 
described in Smith, Sigurdsson, \& Abel (2007), in preparation.  The nonequilibrium 
abundances and cooling rates of H, H$^{+}$, H$^{-}$, He, He$^{+}$, He$^{++}$, H$_{2}$, 
H$_{2}^{+}$, and e$^{-}$ are calculated internally, as in 
\citet{2002Sci...295...93A,1997NewA....2..209A}.  Meanwhile, 
the metal cooling rates are interpolated from large grids of values, precomputed with 
the photoionization software, CLOUDY \citep{1998PASP..110..761F}.  We ignore the cooling 
from dust and focus only on the contribution of gas-phase metals in the optically-thin 
limit.  
Unlike other studies of the formation of the first metal-enriched structures, we do not 
assume the presence of an ionizing UV background.  In our model, the singular pop III star 
that was associated with the dark matter halo in which our stars form has already 
died in a supernova.  We also assume any other Pop III stars are too distant to affect the 
local star-forming region and that QSOs have yet to form.  We use the 
\texttt{coronoal equilibrium} command when 
constructing the cooling data in CLOUDY to simulate a gas where all ionization is 
collisional.  
The metal cooling data 
was created using the Linux cluster, Lion-xo, run by the High Performance Computing Group 
at The Pennsylvania State University.  
As a consequence of our choice to ignore any external radiation, we do not observe the 
fine-structure emission of [C \textsc{ii}] (157.74 $\mu$m) that was reported by 
\citet{2006ApJ...643...26S} to be important.  Instead, cooling from C comes in the form 
of fine-structure lines of [C \textsc{i}] (369.7 $\mu$m, 609.2 $\mu$m).  The cooling from 
[C \textsc{i}] in our study dominates in the same range of densities and temperatures as 
the cooling from [C \textsc{ii}] in \citet{2006ApJ...643...26S}.  
We observe the contributions of the other coolants studied by \citet{2006ApJ...643...26S}, 
[O \textsc{i}], [Si \textsc{ii}], and [Fe \textsc{ii}], to be in agreement with their work.  
In addition, we find that emission from [S \textsc{i}] (25.19 $\mu$m) dominates the 
cooling from metals at n $\sim$ 10$^{7}$ cm$^{-3}$ and T $\sim$ 1--3 $\times$ 10$^{3}$ K.  
The absence of UV radiation in our simulations also allows H$_{2}$ to form, differentiating 
this study from \citet{2001MNRAS.328..969B}.  This allows for a more direct comparison 
between the metal-free and metal-enriched cases.

The simulations are run until one or more dense cores form at the center of the dark 
matter halo and a maximum refinement level of 28 is reached for the first time, giving us 
a dynamic range of greater than 10$^{10}$.  Only the simulation with Z = 10$^{-2}$ Z$\subsun$ 
reached 28 levels of refinement.  The three other simulations were stopped after reaching 27 
refinement levels, since their central densities were already higher than the simulation with 
Z = 10$^{-2}$ Z$\subsun$.  Table 1 summarizes the final state of each simulation, 
where z$_{col}$ is the collapse redshift, l$_{max}$ is the highest level of refinement, 
$n_{max}$ is the maximum gas density within the box, and $\Delta$t$_{col}$ is the time 
difference to collapse from the metal-free simulation.

\section{Results} \label{sec:results}

As can be seen in Table 1, the runs with higher metallicities reach the runaway 
collapse phase faster.  The relationship between metallicity relative to solar and 
$\Delta$t$_{col}$ is well fit by a power-law with index, n $\simeq$ 0.22.  Gas-clouds with 
more metals are able to radiate away their thermal energy more quickly, and thus, collapse 
faster.  
An inverse relation between metallicity and the number of grids and grid-cells exists 
because the low-density, background gas evolves at roughly the same rate in all 
simulations, yet has more time, in the runs with lower metallicities, with which to 
collapse to higher density, requiring additional refinement.

Our simulations, shown in Figure 1, display a qualitative transition in 
behavior between metallicities of 10$^{-4}$ Z$\subsun$ and 10$^{-3}$ Z$\subsun$.  In the 
runs with the highest metallicities (Figure 1C and 1D), the central core is extremely 
asymmetric, and multiple density maxima are clearly visible.  
All four runs display similar large-scale density profiles (Figure 2A).
Radiative cooling from H$_{2}$ becomes extremely inefficient below T $\sim$ 200 K, creating 
the effective temperature floor, visible in Figure 2B for the 
metal-free case \citep{2002Sci...295...93A,2002ApJ...564...23B}.  At n $\simeq$ 10$^{4}$ 
cm$^{-3}$, the rotational levels of H$_{2}$ are populated according to LTE, reducing the 
cooling efficiency and causing the temperature to increase 
\citep{2002Sci...295...93A,2002ApJ...564...23B}.  In the isothermal collapse model of 
\citet{1977ApJ...214..488S}, the accretion rate is proportional to the cube of the sound 
speed.  The increase in temperature leads to an increase in the accretion rate, causing the 
density, and thus, the enclosed mass (Figure 2C), to be slightly higher inside 
the central $\sim$ 0.1 pc 
in the metal-free case.  A similar situation occurs further within for the Z = 10$^{-4}$ 
Z$\subsun$ and, later, the 10$^{-3}$ Z$\subsun$ cases, as the metal cooling is overwhelmed 
by adiabatic compression heating and the temperature begins to rise with density.  The 
presence of metals at the level of 10$^{-4}$ Z$\subsun$ enhances the cooling enough to 
lower the gas temperature to $\sim$ 75 K.  Metallicities greater than 10$^{-3}$ Z$\subsun$ 
provide sufficient cooling to bring the gas down to the temperature of the cosmic microwave 
background, where T$_{CMB}$ $\simeq$ 2.7 K (1 + z).  The gas temperatures are in 
general agreement with the calculations of \citet{2005ApJ...626..627O} that include a CMB 
spectrum at z = 20.  Fragmentation requires that the cooling time be less than the dynamical 
time.  Figure 2D shows that this criterion is essentially never met 
in the zero metallicity case, and only marginally in the Z = 10$^{-4}$ Z$\subsun$ case.  
However, the fragmentation criterion is more than satisfied in the Z = 10$^{-3}$ Z$\subsun$ 
and 10$^{-2}$ Z$\subsun$ cases over a wide mass-range.

In order to locate fragments within our simulations, we employ an 
algorithm, based on \citet{1994ApJ...428..693W}, that works by identifying isolated density 
countours.  Before we search for clumps, we smooth the density field by assigning each 
grid-cell the mass-weighted mean density of the group of cells including itself and its 
neighbors within one cell-width.  This serves to eliminate small density perturbations 
that would be misidentified as clumps by the code.  In order to directly compare the 
fragmentation from each simulation, we limit the search for clumps to the 1 M$\subsun$ of 
gas surrounding the cell with the highest density.  On larger 
scales, all of the runs display a filamentary structure that is qualitatively similar.  
No other region in any of the simulation boxes has collapsed to densities comparable to 
those found within the region where the clump search is performed.  
The results are shown in Figure 3.  A single clump exists in the metal-free and 10$^{-4}$ 
Z$\subsun$ simulations, containing 99.7\% of the total mass within the region of 
interest.  In the simulation with Z = 10$^{-3}$ Z$\subsun$, 91\% of the mass is shared 
between two clumps with 0.52 M$\subsun$ and 0.39 M$\subsun$.  In the same simulation, we also 
find two smaller clumps 0.06 M$\subsun$ and 0.02 M$\subsun$.  Finally, in the Z = 10$^{-2}$ 
Z$\subsun$ simulation, we see two clumps with 0.79 M$\subsun$ and 0.21 M$\subsun$.

\section{Discussion} \label{sec:discussion}

We have shown, through three-dimensional hydrodynamic simulations, that fragmentation 
occurs in collapsing gas with metallicities, Z $\ge$ 10$^{-3}$ Z$\subsun$.  
Our results indicate that star-formation occurs in exactly the same manner at metallicity, 
Z = 10$^{-4}$ Z$\subsun$, as it does at zero metallicity.  The similarities between the 
simulations with metallicities, Z = 10$^{-3}$ Z$\subsun$ and 10$^{-2}$ Z$\subsun$, suggest 
that the transition to low-mass star-formation is complete by 10$^{-3}$ Z$\subsun$, implying 
that the entire transition occurs over only one order of magnitude in metal abundance.  More 
simulations, bracketing the metallicity range, 10$^{-4}$ to 10$^{-3}$ Z$\subsun$, will test 
how abrupt the transition truly is.
We will also explore the effect of non-solar 
abundances on the low metallicity IMF. It has been recently argued that dust cooling 
at high densities (n $\ge$ 10$^{13}$ cm$^{-3}$) can induce fragmentation for metallicities 
as low as 10$^{-6}$ Z$\subsun$ \citet{2006MNRAS.369.1437S}.  In light of the work by 
\citet{2007astro.ph..1395F}, who note the absence of stars with 
D$_{trans}$ $<$ -3.5, where D$_{trans}$ is a measure of the combined logarithmic 
abundance of C and O, it seems unlikely that Z$_{cr}$ is this low.  While the fragmentation 
mode discussed in \citet{2006MNRAS.369.1437S}, and also \citet{2005ApJ...626..627O}, may 
truly exist, it is possible that metal yields from Pop III supernovae overshoot this 
metallicity, for realistic mixing scenarios, 
leaving almost no star-forming regions with such a low concentration of heavy elements.  
Similar to our results, \citet{2005ApJ...626..627O} note that only high-mass fragments are 
produced when Z = 10$^{-4}$ Z$\subsun$.  If Pop III supernovae are able to immediately 
enrich the local universe to Z = 10$^{-4}$ Z$\subsun$, the high-density dust cooling fragmentation 
mode would be skipped altogether, and the high-mass stars that formed via the mode observed 
at 10$^{-4}$ Z$\subsun$ would leave no record in the search for low-metallicity stars in 
the local universe.

We have limited the search for fragments to the dense 1 M$\subsun$ core at the center of each 
simulation.  Within this region, it is unlikely that any more fragments will form in any of the 
simulations.  In all of the cases presented, the cooling has begun to be overwhelmed by 
compression heating such that the central temperature is now increasing with increasing density, 
which was indicated by \citet{2005MNRAS.359..211L} to be the end of hierarchical fragmentation.  
Fragmentation may continue in the surrounding lower density gas in the cases of Z = 10$^{-3}$ 
Z$\subsun$ and 10$^{-2}$ Z$\subsun$.  The final stellar masses of these objects will also 
be affected interaction and accretion that will occur in later stages of evolution.  
In the two lowest metallicity cases, the gas immediately 
surrounding the central core evolves slowly enough that it will not have sufficient time to 
reach high densities before the UV radiation from the central, massive star dissociates all of 
the H$_{2}$.  As was shown by \citet{2001MNRAS.328..969B}, clouds with metallicities, Z 
$\le$ 10$^{-4}$ Z$\subsun$ are unable to collapse without the aid of H$_{2}$ cooling.

In the two simulations in which significant fragmentation is observed, Z = 10$^{-3}$ and 
10$^{-2}$ Z$\subsun$, the gas is able to cool rapidly to the temperature of the CMB.  
\citet{2005ApJ...629..615W} predict that the rate of Pop III supernovae 
peaks at a redshift, z $\sim$ 20, and then drops off sharply, implying that metal production 
from Pop III stars is effectively finished at this point.  In this epoch, the characteristic 
mass-scale for metal-enriched star formation will be regulated by the CMB, as is predicted in 
\citet{2003Natur.425..812B}.  Thus, the first Pop II stars will be considerably more 
massive, on average, 
than stars observed today, as was suggested by \citet{1998MNRAS.301..569L}.  Observations of 
low-mass prestellar cores in the local universe reveal them to have 
temperatures of about 8.5 K \citep{1999ARA&A..37..311E}, implying that the IMF may not 
become completely 'normal' until z $<$ 3 when the CMB fell below this temperature. 

\acknowledgments
We thank Tom Abel, Greg Bryan, Mike Norman, Brian O'Shea, and Matt Turk for useful 
discussions.  BDS also thanks Michael Kuhlen for providing an update to some useful 
analysis tools.  We are also very grateful for insightful comments from an anonymous referee.
This work was made possible by Hubble Space Telescope Theory 
Grant HST-AR-10978.01, and an allocation from the San Diego Supercomputing Center.

\clearpage

\begin{tabular}{ccccccc}
\multicolumn{7}{c}
	{\rule[-3mm]{0mm}{8mm}Table 1}\\
\multicolumn{7}{c}
	{\rule[-3mm]{0mm}{8mm}Simulation Final States}\\
	\hline
	\hline
Z (Z$\subsun$) &
z$_{col}$      &
l$_{max}$      &
Grids          &
Cells          &
$n_{max}$ (cm$^{-3}$) &
$\Delta$t$_{col}$ (Myr)\\
	\hline
0           & 18.231519 & 27 & 8469 & 4.82 $\times 10^{7}$ & 4.11 $\times 10^{13}$ & -\\
10$^{-4}$   & 18.838816 & 27 & 8060 & 4.64 $\times 10^{7}$ & 3.90 $\times 10^{13}$ & 9.19\\
10$^{-3}$   & 19.336557 & 27 & 7911 & 4.56 $\times 10^{7}$ & 1.65 $\times 10^{13}$ & 16.21\\
10$^{-2}$   & 20.032518 & 28 & 7521 & 4.42 $\times 10^{7}$ & 1.50 $\times 10^{13}$ & 25.33\\
	\hline
\end{tabular}

\clearpage

\begin{figure}
  \plotone{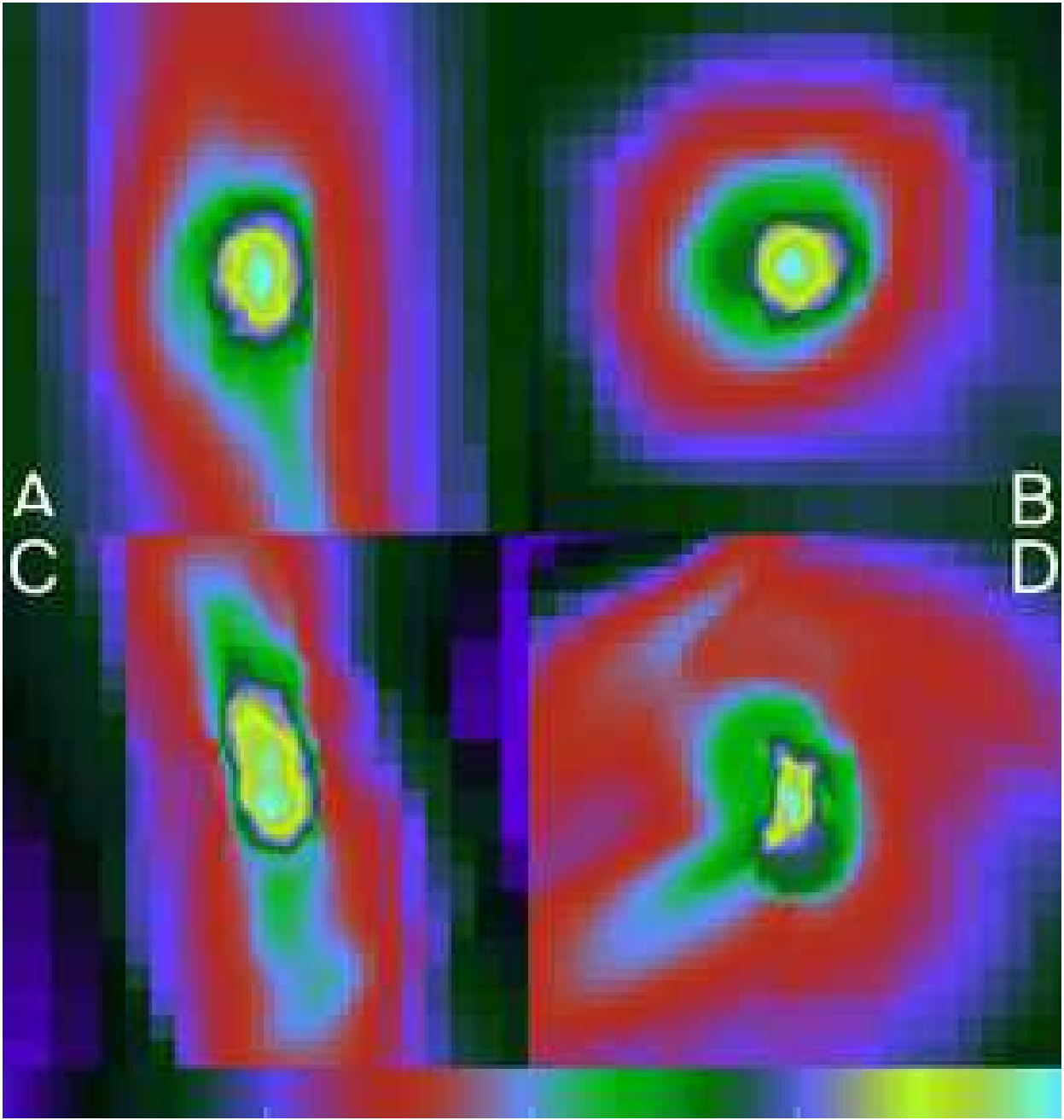}
  \caption{Slices through gas density for the final output of simulations with Z = 0 
    (A), 10$^{-4}$ Z$\subsun$ (B), 10$^{-3}$ Z$\subsun$ (C), and 
    10$^{-2}$ Z$\subsun$ (D).  Each slice intersects the grid-cell with the 
    highest gas density and has a width of 2 $\times$ 10$^{-8}$ of the computation box, 
    corresponding to a proper size of $\sim$4 $\times$ 10$^{-4}$ pc (84 AU).  The color-bar 
    at bottom ascends logarithmically, from left to right, spanning exactly four orders of 
    magnitude in density.}
\end{figure}

\clearpage

\begin{figure}
  \includegraphics[width=6in,height=6in,angle=-90]{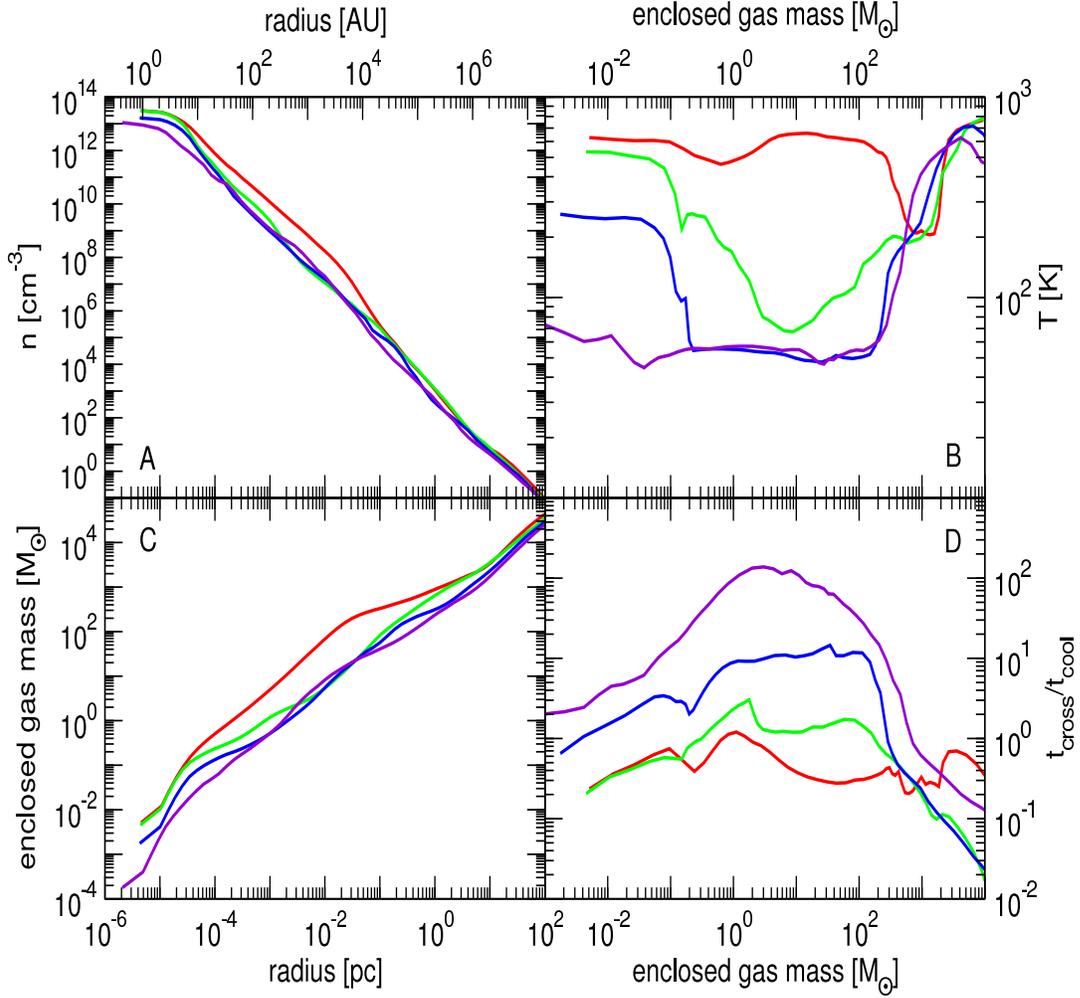}
  \caption{Radially averaged, mass-weighted quantities for the final output each 
    simulation: Z = 0 (red), 10$^{-4}$ Z$\subsun$ (green), 10$^{-3}$ Z$\subsun$ (blue), 
    and 10$^{-2}$ Z$\subsun$ (purple).  
    A: Number density vs. radius.  B: Temperature vs. enclosed mass.  C: Enclosed gas 
    mass vs. radius.  D: Ratio of crossing time to cooling time vs. enclosed mass.  
    The classical criterion for fragmentation is met when the ratio of the crossing 
    time to the cooling time is greater than 1.}
\end{figure}

\clearpage

\begin{figure}
  \includegraphics[width=6in,height=6in,angle=-90]{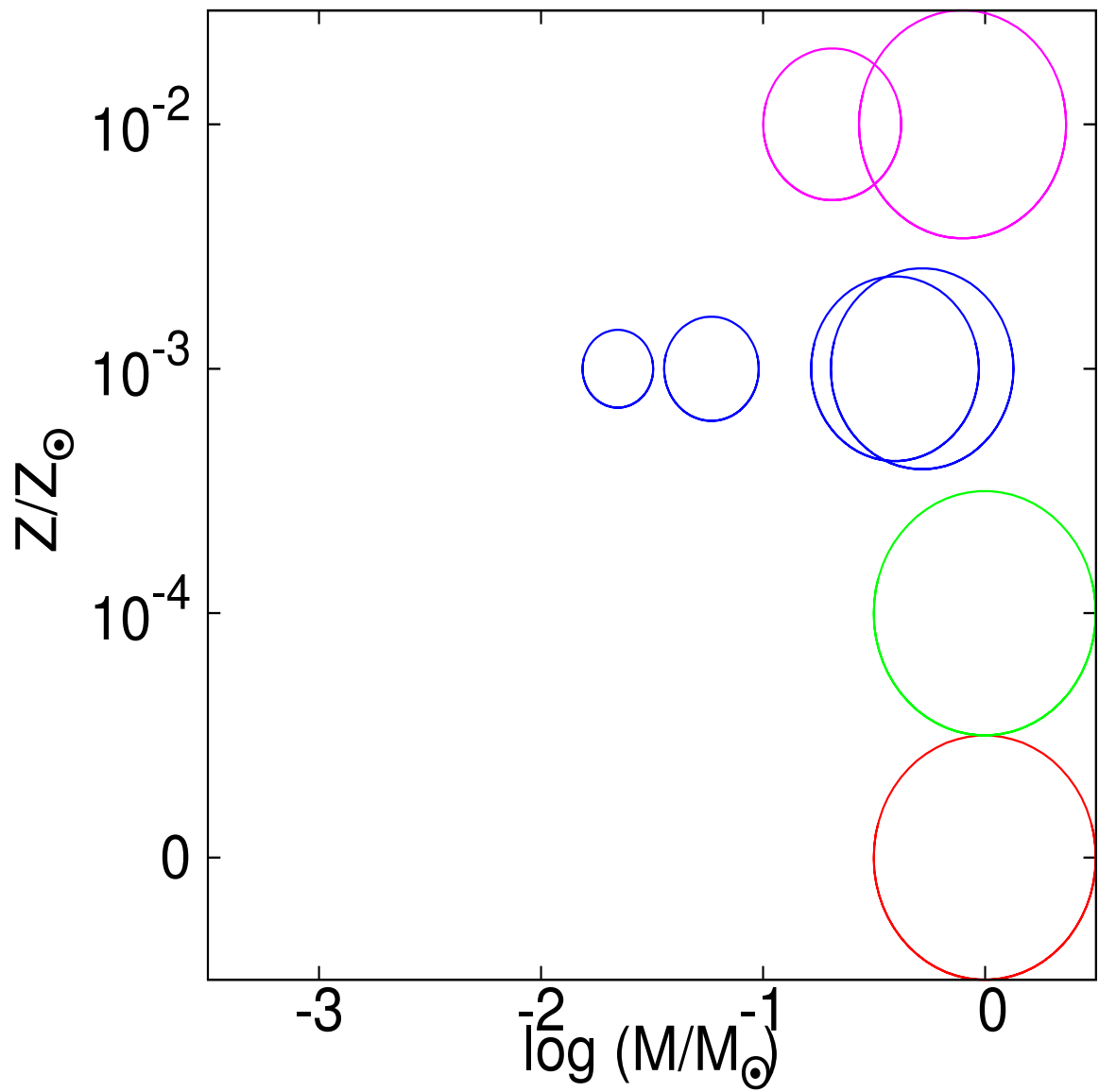}
  \caption{Masses of clumps found within the final output of each simulation.  The 
    location on the x and y axes corresponds to the log of the clump mass and the 
    metallicity of the simulation.  Colors are the same as in Figure 2.  
    The radii of the circles are proportional to the masses of the clumps they represent.  
    A factor of 10 in mass is equivalent to a factor of 2 in radius.  The search for 
    clumps is limited to the 1 M$\subsun$ surrounding the grid cell with the highest gas 
    density.  Only clumps with at least 1000 cells are plotted.}
\end{figure}

\end{document}